\documentclass[epj]{webofc}
\usepackage[utf8]{inputenc}
\usepackage[varg]{txfonts}   
\usepackage{booktabs}
\usepackage{xcolor}
\definecolor{darkred}{rgb}{0.4,0.0,0.0}
\definecolor{darkgreen}{rgb}{0.0,0.4,0.0}
\definecolor{darkblue}{rgb}{0.0,0.0,0.4}
\usepackage[bookmarks,linktocpage,colorlinks,
    linkcolor = darkred,
    urlcolor  = darkblue,
    citecolor = darkgreen]{hyperref}
\usepackage{subfigure}
\wocname{EPJ Web of Conferences}
\woctitle{Lattice2017}
\begin{document}
%
\selectlanguage{english}
\title{%
Comparison of different source calculations in two-nucleon channel at large quark mass
}
\author{%
\firstname{Takeshi} \lastname{Yamazaki}\inst{1,2,3}\fnsep\thanks{Speaker, \email{yamazaki@het.ph.tsukuba.ac.jp}} \and
\firstname{Ken-ichi} \lastname{Ishikawa}\inst{4} \and
\firstname{Yoshinobu}  \lastname{Kuramashi}\inst{2,3}
\ for PACS Collaboration
}
\institute{%
Faculty of Pure and Applied Sciences,
University of Tsukuba, Tsukuba, Ibaraki 305-8571, Japan
\and
Center for Computational Sciences, University of Tsukuba,
Tsukuba, Ibaraki 305-8577, Japan
\and
RIKEN Advanced Institute for Computational Science,
Kobe, Hyogo 650-0047, Japan
\and
Department of Physics, Hiroshima University, Higashi-Hiroshima, Hiroshima 739-8526, Japan
}
\abstract{
We investigate a systematic error coming from higher excited state 
contributions in the energy shift of light nucleus in the two-nucleon
channel by comparing 
two different source calculations with the exponential and 
wall sources. Since it is hard to obtain a clear signal of the wall source
correlation function in a plateau region, we employ a large quark mass
as the pion mass is 0.8 GeV in quenched QCD. We discuss the systematic
error in the spin-triplet channel of the two-nucleon system, and 
the volume dependence of the energy shift.
}
\maketitle
\section{Introduction}\label{sec:intro}

We carried out an exploratory study of the direct calculation of 
the binding energy of the light nuclei with the atomic mass number 
less than or equal to four
in quenched lattice QCD~\cite{Yamazaki:2009ua,Yamazaki:2011nd}.
These studies were followed by several calculations~\cite{Beane:2011iw,Beane:2012vq,Yamazaki:2012hi,Berkowitz:2015eaa,Yamazaki:2015asa,Orginos:2015aya,Wagman:2017tmp}.
All the recent calculations at $m_\pi > 0.3$ GeV,
which were obtained from the calculations 
with the exponential or gaussian source,
suggest the existence of a bound state in the two-nucleon channels.

HALQCD~\cite{Iritani:2016jie} suggested that
there is a sizable systematic error in the energy shift
in the two-nucleon channels
obtained from the ratio of the correlation functions.
They compared the two results with the exponential and wall sources,
and found discrepancies in the effective energy shifts.
However, it is well known that the wall source needs the longest
temporal extent to obtain a plateau even in the single nucleon mass.
In this comparison a high precision calculation is necessary.

The purpose of this work is to investigate the systematic error
coming from excited states by comparing the exponential and wall source
calculations in the spin-triplet two-nucleon channel
in a high precision calculation using 
a large quark mass of $m_\pi = 0.8$ GeV in the quenched approximation.
To determine a plateau of the ratio of the correlation functions,
we focus on an important condition in the direct calculation,
which will be explained below, though it is trivial in lattice QCD calculation.
The results in this report are the updated ones from 
the last conference~\cite{Yamazaki:2017euu}.
All the results in this report are preliminary.

\section{Important condition of direct calculation}\label{sec:condition}

In the direct calculation~\cite{Yamazaki:2009ua,Yamazaki:2011nd,Beane:2011iw,Beane:2012vq,Yamazaki:2012hi,Berkowitz:2015eaa,Yamazaki:2015asa,Orginos:2015aya,Wagman:2017tmp} of the two-nucleon channel, 
the energy shift $\Delta E_{NN} = 2 m_N - E_{NN}$
with the nucleon mass $m_N$ and two-nucleon ground state energy
$E_{NN}$ is determined from
a plateau region of the ratio of the correlation functions 
$R(t) = C_{NN}(t)/C^2_N(t)$ with the two-nucleon correlation function 
$C_{NN}(t)$ in the spin-triplet
channel and the single nucleon correlation function $C_N(t)$. 
An important condition of this determination is 
that $\Delta E_{NN}$ should be determined 
in a region where both $C_{NN}(t)$ and $C^2_N(t)$ have each plateau.
It means that it is not enough to determine a plateau region 
from only $R(t)$,
but we need to investigate plateaus for $C_{NN}(t)$ and $C^2_N(t)$.

If one chooses a plateau from only $R(t)$,
it might cause an incorrect determination of $\Delta E_{NN}$,
as discussed in the later sections.
For example, when statistics is not enough, $R(t)$ using the wall source 
has a plateau like behavior in early $t$ region, where 
$C_{NN}(t)$ and $C^2_N(t)$ do not have plateaus.

In the following sections, we shall call the minimum $t$ of the plateau region
for $C_{NN}(t)$, $C_N(t)$, and $R(t)$ as $t_N^S$, $t_{NN}^S$, and $t_R^S$,
respectively, using the source $S = E$ (exponential) or $W$ (wall).

\section{Simulation parameters}\label{sec:params}

We calculate the two-nucleon correlation function 
$C_{NN}(t)$ in the spin-triplet
channel as well as the single nucleon correlation function $C_N(t)$ 
in the quenched approximation.
In this calculation,
we employ Iwasaki gauge action
at $\beta = 2.416$, corresponding to $a = 0.128$ fm~\cite{AliKhan:2001tx}.
The quark propagators are calculated with a tad-pole improved Wilson
action with $c_{\rm SW} = 1.378$ at $\kappa_{ud} = 0.13482$ corresponding
to $m_\pi = 0.8$ GeV and $m_N =1.62$ GeV.
The actions and parameters are the same as in 
our previous works~\cite{Yamazaki:2009ua,Yamazaki:2011nd}.
The temporal lattice size is fixed to 64, while the spatial size $L$ is
chosen to be 16, 20, and 32.

In order to compare results with different source operators,
we employ the exponential and wall sources.
The exponential source at the time slice $t$ is defined by
\begin{equation}
q^\prime({\bf x},t) = q({\bf x},t) + A \sum_{{\bf y} \ne {\bf x}} 
\exp(-B |{\bf y}-{\bf x}|) q({\bf y},t),
\end{equation}
where $q({\bf x},t)$ is the local quark field.
The parameters $A$ and $B$ are chosen to obtain an early plateau of
the effective nucleon mass on each volume.
At the sink time slice, each nucleon operator is projected to
zero momentum using the local quark field as in our previous
calculations~\cite{Yamazaki:2011nd,Yamazaki:2012hi,Yamazaki:2015asa}.
The number of the measurement of the correlation functions is
tabulated in Table~\ref{tab:nmeas}.

\begin{table}[!h]
\hfil
\begin{tabular}{|c|c|c|c|}\hline
$L$  & 16 & 20 & 32 \\\hline
Exp  & 6,272,000 & 5,504,000 & 4,736,000 \\\hline
Wall & 8,307,200 & 8,960,000 & 4,473,600 \\\hline
\end{tabular}
\caption{
Numbers of the measurement on each $L$ with the exponential (Exp)
and wall sources.
\label{tab:nmeas}}
\end{table}

\section{Results}\label{sec:results}

In this section we present the results for twice the effective nucleon
mass $2 m_N^{\rm eff}$, the effective two-nucleon energy $E_{NN}^{\rm eff}$,
and the effective energy shift $\Delta E_{NN}^{\rm eff}$
evaluated from $C_N(t)$, $C_{NN}(t)$, and $R(t)$, respectively, on
each volume.
The volume dependence of $\Delta E_{NN}$ is also presented.

\subsection{$L=20$}

First we present the results of 
$2 m_N^{\rm eff}$, $E_{NN}^{\rm eff}$, and $\Delta E_{NN}^{\rm eff}$
in both the exponential and wall sources
on $L=20$ as a typical result.

\begin{figure}[!t]
\hspace*{-2mm}
\begin{tabular}{cc}
\includegraphics*[angle=0,width=0.48\textwidth]{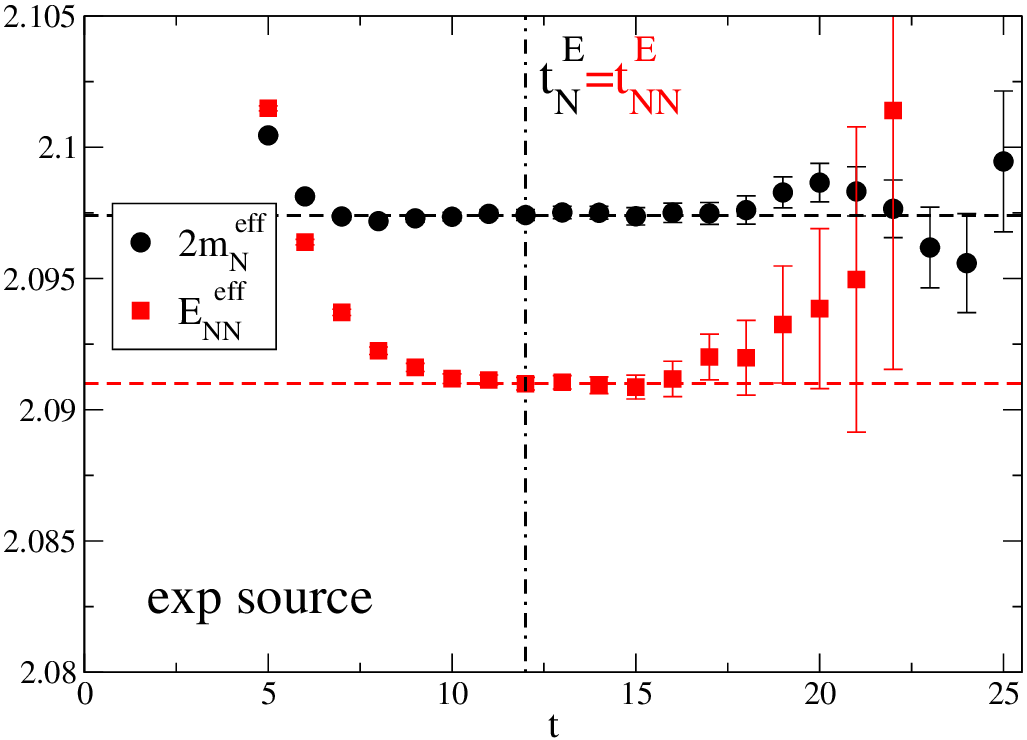}
&
\includegraphics*[angle=0,width=0.48\textwidth]{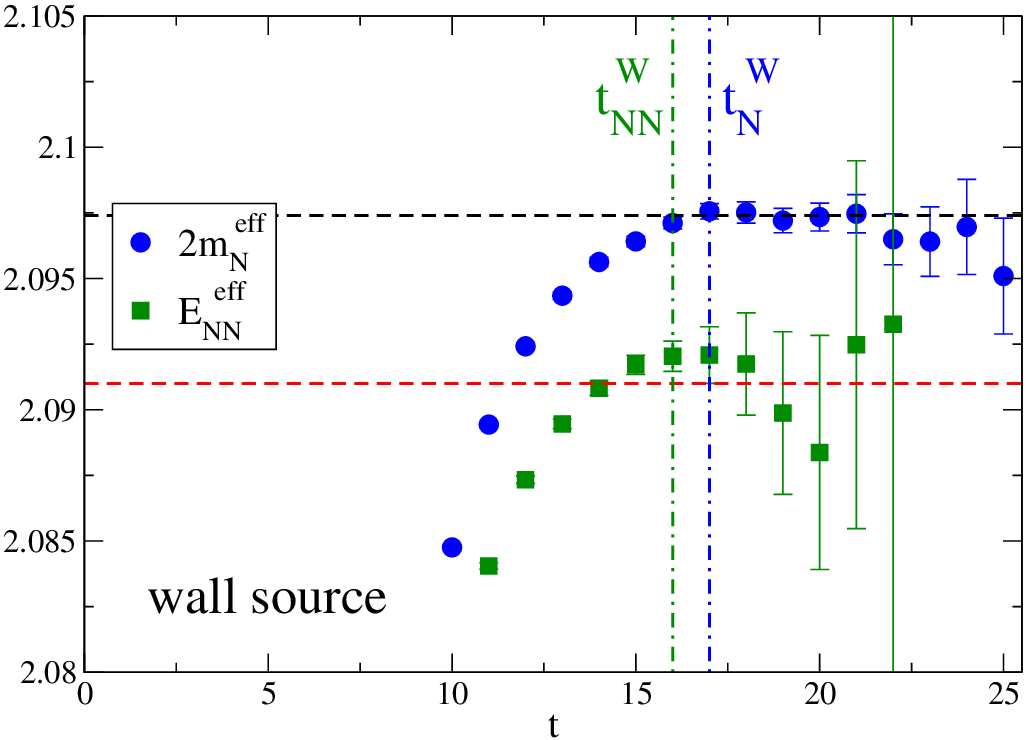}
\end{tabular}
\caption{
Effective twice nucleon mass $2m_N^{\rm eff}$ (circle) and 
effective two-nucleon energy $E_{NN}^{\rm eff}$ (square) 
in the spin-triplet channel
using the exponential (left panel) and wall (right panel) sources
on $L=20$.
The vertical dot-dashed lines denote $t_N^S$ and $t_{NN}^S$ for $S=E,W$
explained in the text.
The horizontal dashed lines express the values of each plateau in the 
exponential source in both the panels.
\label{fig:L20:comp}
}
\end{figure}

Figure~\ref{fig:L20:comp} shows the results for
$2 m_N^{\rm eff}$ and $E_{NN}^{\rm eff}$ using
the exponential (left panel) and wall (right panel) sources.
The results of the exponential source have plateaus,
which start from $t=12$ as denoted by vertical dot-dashed line
in the left panel.
It means that $t_N^E = t_{NN}^E = 12$ in this case.
The horizontal dashed lines in black and red
represent the values of the plateaus
for $2 m_N^{\rm eff}$ and $E_{NN}^{\rm eff}$, respectively.

The wall source results need longer $t$ than the exponential source
to have plateaus as shown in the right panel of Fig.~\ref{fig:L20:comp}.
We determine $t_N^W = 17$ and $t_{NN}^W = 16$ from each plateau region,
which are expressed by vertical dot-dashed lines.
The same horizontal lines as in the left panel are shown in the right panel.
Those lines are in good agreement with each plateau.

\begin{figure}[!t]
\hfil
\includegraphics*[angle=0,width=0.49\textwidth]{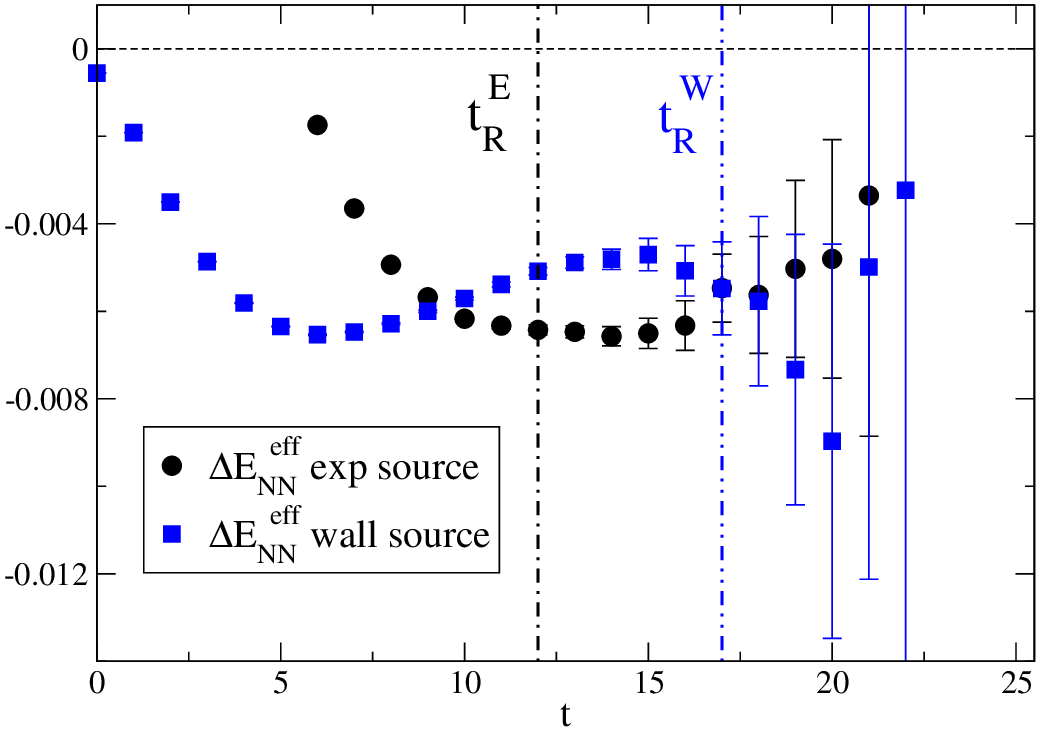}
\caption{
Effective energy shift 
$\Delta E_{NN}^{\rm eff} = 2m_N^{\rm eff}-E_{NN}^{\rm eff}$
using the exponential (circle) and wall (square) sources
on $L=20$.
The vertical dot-dashed and dot-dot-dashed lines express
$t_R^E$ and $t_R^W$, respectively, as explained in the text.
\label{fig:L20:comp:dE}
}
\end{figure}

As discussed in Sec.~\ref{sec:condition},
$t_R^S$, the minimum $t$ of 
the plateau region of $\Delta E_{NN}^{\rm eff}$,
should be larger or equal to $t_N^S$ and $t_{NN}^S$.
Thus, $t_R^E = t_N^E = t_{NN}^E$, and $t_R^W = t_N^W$, in this case.
Figure~\ref{fig:L20:comp:dE} presents that 
$\Delta E_{NN}^{\rm eff}$ of the exponential source has a reasonable
plateau after $t_R^E$.
On the other hand, the result of the wall source has a non-monotonic $t$
dependence in $t < 15$.

One might choose $t_R^W \sim 14$,
if it is determined from only the wall source data
in Fig.~\ref{fig:L20:comp:dE}.
Moreover, if the statistics is much smaller than the current calculation,
the data around $t = 5$ would be also regarded as a plateau.
However, the data in the small $t$ region contain excited state contributions
as shown in the right panel of Fig.~\ref{fig:L20:comp}.
It suggests that it is easy to mistake the plateau region, when
it is determined from only $\Delta E_{NN}^{\rm eff}$,
especially in the case where $\Delta E_{NN}^{\rm eff}$ has a non-monotonic
$t$ dependence, like the wall source data in the current study.

In the wall source, while the data in $t \ge t_R^W$ has the large error,
it agrees with the plateau value of the exponential source within the error.

\subsection{$L=16$}

\begin{figure}[!t]
\hspace*{-2mm}
\begin{tabular}{cc}
\includegraphics*[angle=0,width=0.48\textwidth]{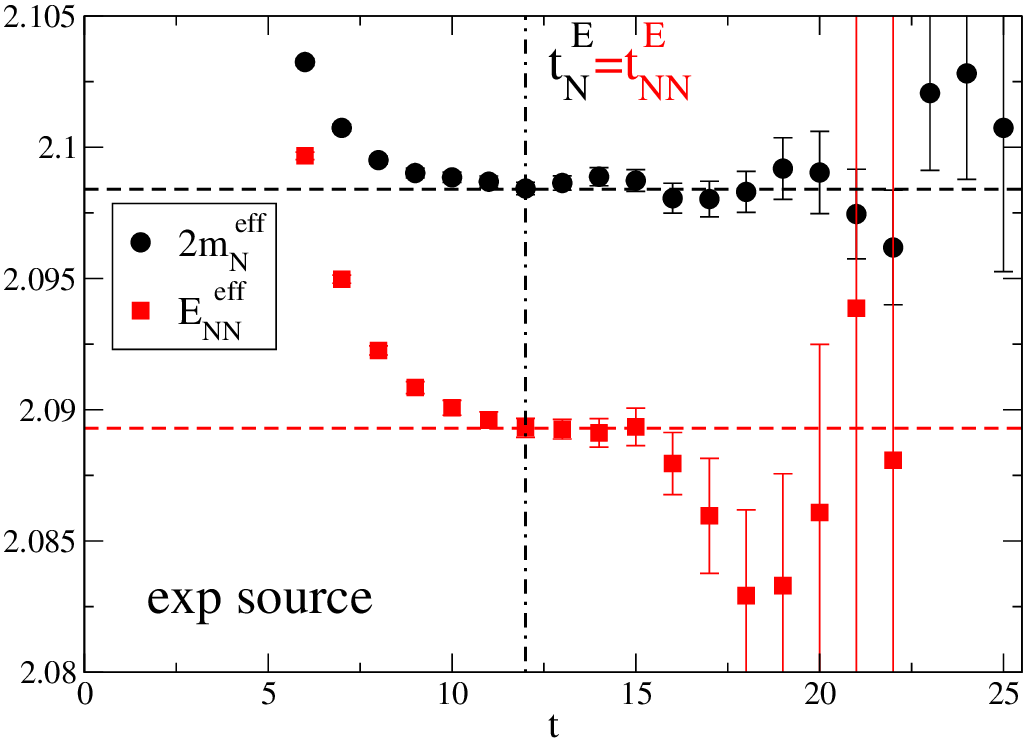}
&
\includegraphics*[angle=0,width=0.48\textwidth]{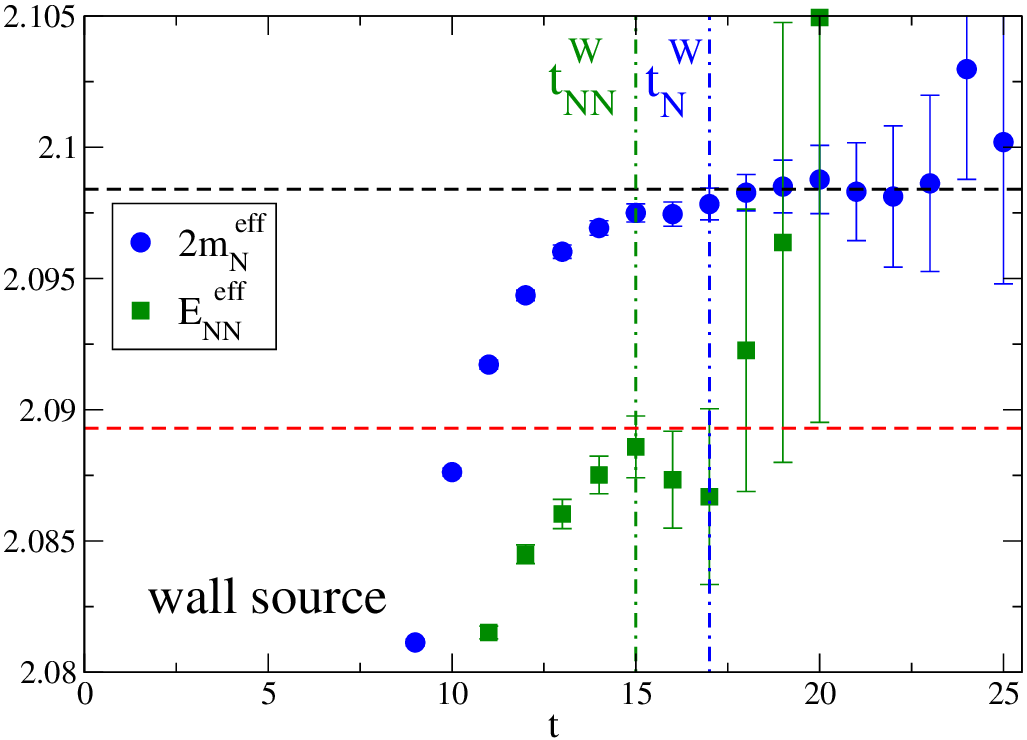}
\end{tabular}
\caption{
The same figures as Fig.~\ref{fig:L20:comp}, but in the $L=16$ case.
\label{fig:L16:comp}
}
\end{figure}

The results for $2 m_N^{\rm eff}$ and $E_{NN}^{\rm eff}$ using
the exponential and wall sources are plotted in the left and
right panels of Fig.~\ref{fig:L16:comp}, respectively.
The results are similar to the ones on $L=20$ in the previous subsection.
The data of the exponential source have plateaus, which start from
$t = 12$, and the ones of the wall source need longer $t$ to
have plateaus.
It is noted that comparing with the results on $L=20$ and 16
we observe 0.02\% finite volume effect in $m_N$
on this volume of the spatial extent $2.0$ fm.

\begin{figure}[!t]
\hfil
\includegraphics*[angle=0,width=0.49\textwidth]{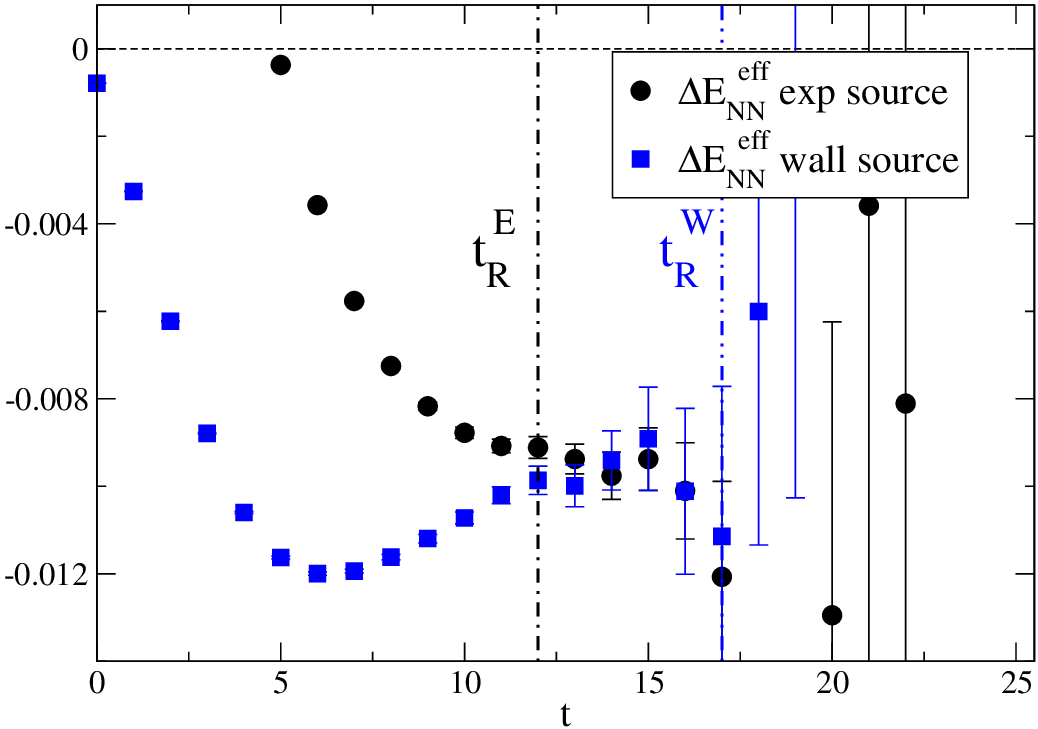}
\caption{
The same figures as Fig.~\ref{fig:L20:comp:dE}, but in the $L=16$ case.
\label{fig:L16:comp:dE}
}
\end{figure}

The results of $\Delta E_{NN}^{\rm eff}$ are 
shown in Fig.~\ref{fig:L16:comp:dE}.
The exponential source has a reasonable plateau after $t_R^E = t_N^E = t_{NN}^E$
as in the $L=20$ case.
The $t$ dependence of the wall source in the smaller $t$ region
becomes larger as the volume decreases comparing with the result
in Fig.~\ref{fig:L20:comp:dE}.
While the error of the wall source is large after $t_R^W = t_N^W$, 
the data is consistent with the plateau value of
the exponential source.

It is noted that on $L=16$
the consistent results with the exponential and wall sources
are not obtained even in $2 m_N^{\rm eff}$,
when the number of the measurement of 
the wall source is half of the current calculation.
This suggests that a huge statistics is necessary to obtain
statistically stable result from the wall source even in
the single nucleon mass.

\subsection{$L=32$}

\begin{figure}[!t]
\hspace*{-2mm}
\begin{tabular}{cc}
\includegraphics*[angle=0,width=0.48\textwidth]{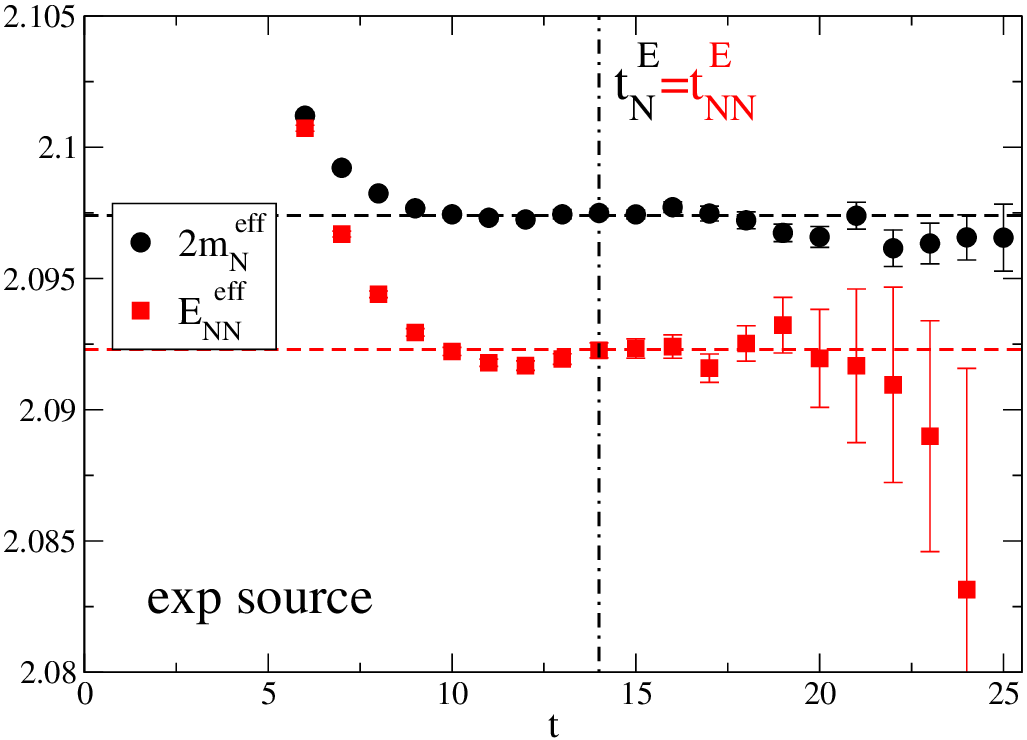}
&
\includegraphics*[angle=0,width=0.48\textwidth]{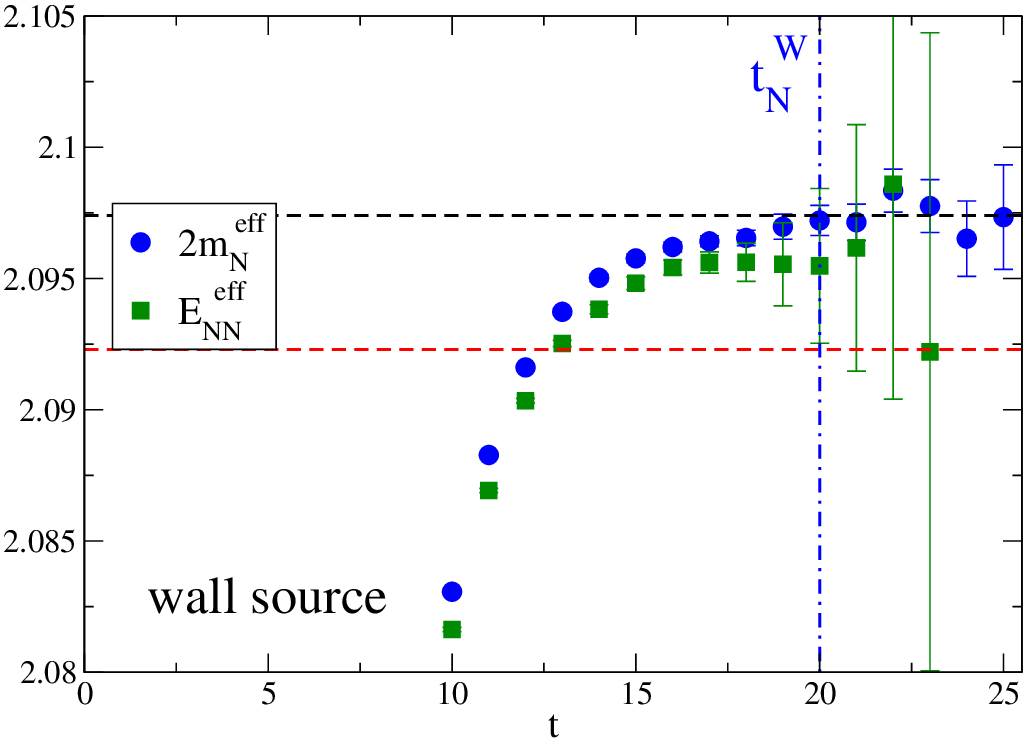}
\end{tabular}
\caption{
The same figures as Fig.~\ref{fig:L20:comp}, but in the $L=32$ case.
\label{fig:L32:comp}
}
\end{figure}

The left panel of Fig.~\ref{fig:L32:comp} presents that
the results for $2 m_N^{\rm eff}$ and $E_{NN}^{\rm eff}$ with
the exponential source are similar to the ones in
the $L=16$ and 20 cases.

On the other hand, the wall source results look different from
the ones on the other volumes.
The result of $E_{NN}^{\rm eff}$ with the wall source in $t \le 20$
is larger than the one with the exponential source
represented by the red dashed line.
One of the reasons is that
the contribution of the two-nucleon scattering
state with almost zero relative momentum, 
which corresponds to the first excited state
in this system, becomes relatively larger than the one of the
ground state in $C_{NN}(t)$ with the wall source as the volume increases.
Another reason is that the energy of the first excited state is
larger than $2m_N$ in this system, where one bound state exists~\cite{Yamazaki:2011nd}.
Thus, it is harder to obtain the same plateau 
as the one of the exponential source,
expressed by the red dashed line in the right panel of Fig.~\ref{fig:L32:comp},
from the wall source as the volume increases.
From the data, we cannot determine $t_{NN}$ of the wall source.
In the following analysis, it is assumed that $t_N^W = t_{NN}^W$ in 
the wall source result.

From the above reasons, 
it is expected that on much larger volumes than the current calculation
$E_{NN}^{\rm eff}$ of the wall source would
become larger than $2 m_N^{\rm eff}$ in a large $t$ region.
Then, it would go down to agree with the plateau value of the exponential
source in much larger $t$ region.

\begin{figure}[!t]
\hfil
\includegraphics*[angle=0,width=0.49\textwidth]{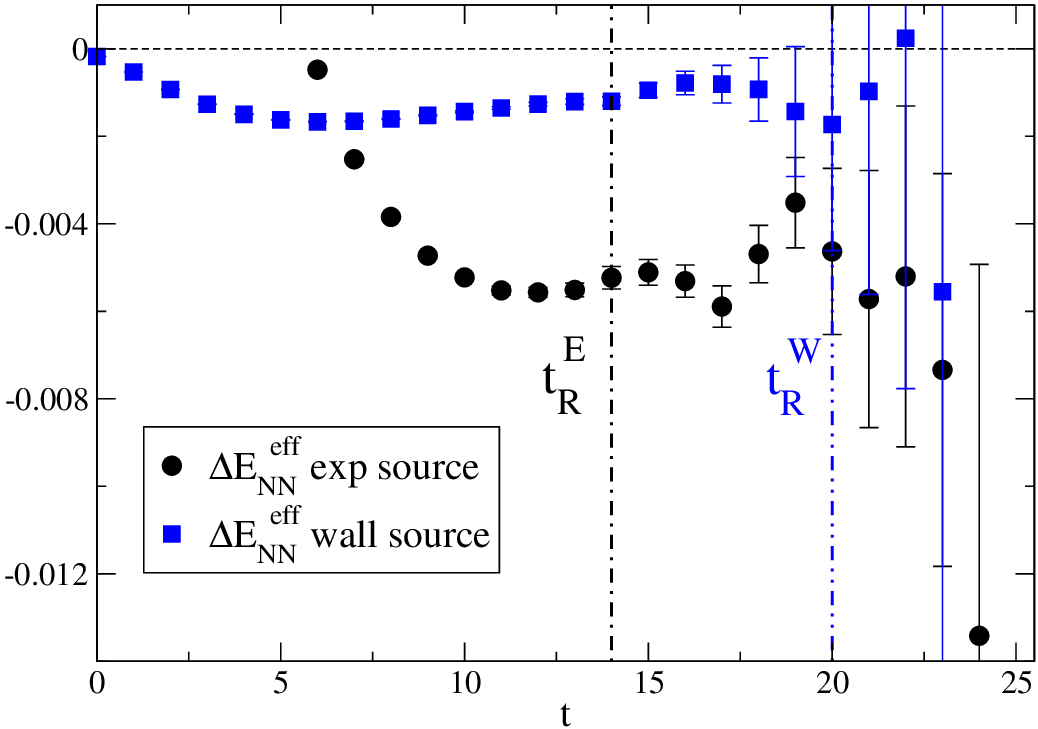}
\caption{
The same figures as Fig.~\ref{fig:L20:comp:dE}, but in the $L=32$ case.
\label{fig:L32:comp:dE}
}
\end{figure}

Figure~\ref{fig:L32:comp:dE} shows the results of $\Delta E_{NN}^{\rm eff}$
with both the sources.
It is surprising that the wall source result has
a mild $t$ dependence in the small $t$ region, although the data for 
$2 m_N^{\rm eff}$ and $E_{NN}^{\rm eff}$ largely depend on $t$
in the same region as shown in the right panel of Fig.~\ref{fig:L32:comp}.
If a plateau of $\Delta E_{NN}^{\rm eff}$ is determined 
from only the wall source data in smaller statistics than
the present calculation, 
one might choose much smaller $t$ region as a plateau than $t_R^W$.

While it is not as good as the smaller volumes,
we observe a plateau after $t_R^E$ in the exponential source result
on this volume.
Although the wall source result has large error after $t_R^W$,
it is not inconsistent with the plateau of the exponential source.
We expect that the plateau of $E_{NN}^{\rm eff}$ with the wall source
is obtained in a region of the much larger $t$ than the smaller volumes.
In order to confirm this expectation,
it is an important future work to observe clear signal of the wall source
after $t_R^W$.

From the comparisons including the ones in the smaller volumes,
we conclude that the results using the exponential and wall source
are consistent with each other in each plateau region.
Thus, contaminations of excited states in $\Delta E_{NN}^{\rm eff}$
obtained from the plateau region
are negligible in our calculation.

\subsection{Volume dependence}

\begin{figure}[!t]
\hfil
\includegraphics*[angle=0,width=0.49\textwidth]{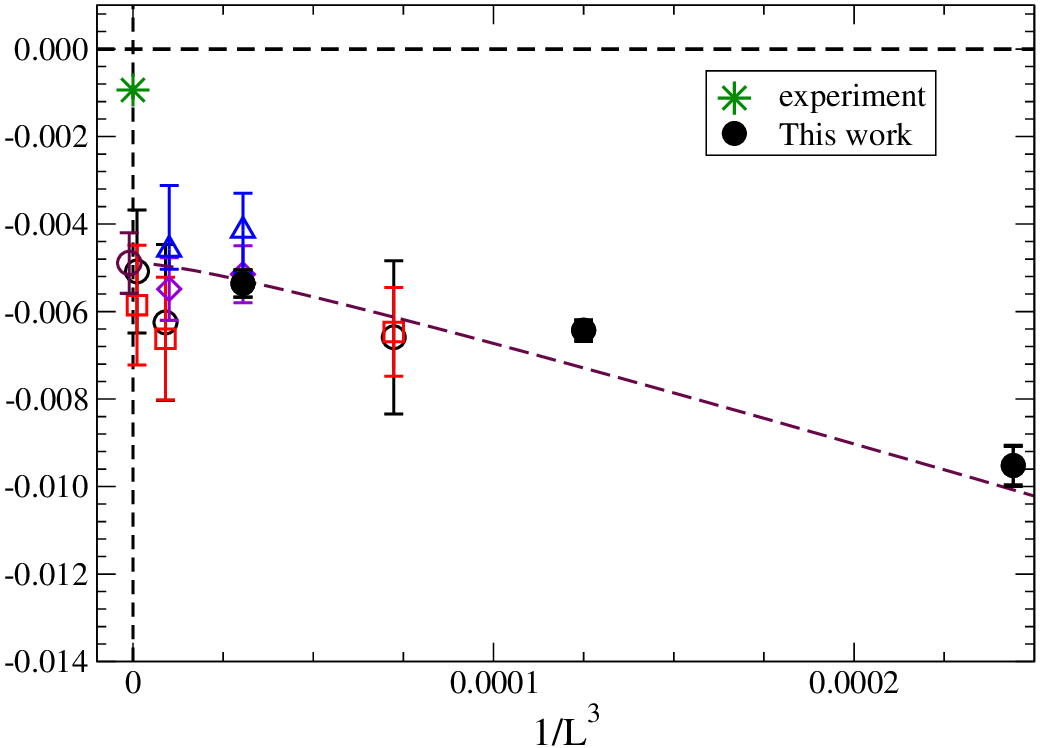}
\caption{
\protect
Volume dependence of the energy shift $\Delta E_{NN}$
using the exponential source.
The horizontal axis is one over volume.
The filled circles denote
$\Delta E_{NN}$ on $L=16, 20, 32$ in the current calculation.
The open symbols and dashed curve are the results in 
our previous work~\cite{Yamazaki:2011nd}.
The star symbol expresses the experimental value of
the deuteron binding energy.
\label{fig:nf0:vdep}
}
\end{figure}

The result of $\Delta E_{NN}$ on the three volumes with
the exponential source are plotted 
in Fig.~\ref{fig:nf0:vdep} together with 
our previous result~\cite{Yamazaki:2011nd}.
We neglect the wall source data in the following due to 
the much larger error.
The result of the current calculation denoted by the filled circle
has much smaller statistical error, and 
is reasonably consistent with the fit curve using the previous data,
so that the result indicates that the existence of a bound state in this system.

Recently HALQCD Collaboration suggested that the volume dependence
of $\Delta E_{NN}$ obtained from the direct calculation is
too small comparing to the one expected from 
the effective range expansion~\cite{Iritani:2017rlk}.
However, this argument is assumed that
the effective range expansion is valid in $p^2 < 0$ region
in the continuum theory, and there is no finite volume effect
in the two-nucleon interaction.
In the comparison between the expectation in the ideal case
and the lattice data,
there could be several sources of systematic errors,
such as finite lattice spacing and finite volume effects,
which may deform the two-nucleon interaction.
In order to understand the current situation,
it is an important future work to investigate such systematic errors
in the $\Delta E_{NN}$ calculation.

It is noted that even if there is a finite volume effect in $\Delta E_{NN}$, 
which cannot be treated by
the finite volume method~\cite{Luscher:1986pf,Luscher:1990ux},
we consider that the signal of the existence of the bound state is 
meaningful in our calculation,
because we discuss the existence in the infinite volume limit, so that
our result does not contain the finite volume effect.

\section{Summary}

We have carried out the high precision calculation of 
the spin-triplet two-nucleon channel at the large quark mass,
corresponding to $m_\pi = 0.8$ GeV in the quenched approximation
to investigate a systematic error of $\Delta E_{NN}$
coming from excited states
by comparing the results with the two different source calculations
using the exponential and wall sources on the three volumes.
Though it might be a trivial, we discuss the important condition
to calculate $\Delta E_{NN}$.
When the condition is satisfied,
the two sources give the consistent results of $\Delta E_{NN}^{\rm eff}$
in each plateau region, while the wall source data has the large error
due to the late plateau.
From this comparison, we have concluded that the systematic error
from higher excited states is negligible in our calculation.

There are several important future works, such as 
comparing the current result with the one obtained from
the generalized eigenvalue problem~\cite{Luscher:1990c},
and
investigations of systematic errors in $\Delta E_{NN}$.
It is also an important future work 
to clarify the qualitative difference between 
the direct calculation and HALQCD method in the point of view of
the definitions of the scattering amplitude in 
quantum field theory and quantum mechanics~\cite{Yamazaki:2017gjl}.

\section*{Acknowledgements}
Numerical calculations for the present work have been carried out on 
the FX10 supercomputer system at Information Technology Center of 
the University of Tokyo, on the COMA cluster system under 
the ``Interdisciplinary Computational Science Program'' of 
Center for Computational Science at University of Tsukuba, 
on the Oakforest-PACS system of Joint Center for 
Advanced High Performance Computing,
on the computer facilities of the Research Institute for 
Information Technology of Kyushu University, and on
the FX100 and CX400 supercomputer systems at the Information Technology Center 
of Nagoya University.
This research used computational resources of the HPCI system provided by 
Information Technology Center of the University of Tokyo through 
the HPCI System Research Project (Project ID: hp160125). 
We thank the colleagues in the PACS Collaboration for providing us 
the code used in this work. This work is supported in part by Grants-in-Aid 
for Scientific Research from the Ministry of Education, Culture, Sports, 
Science and Technology (No. 16H06002).


\bibliography{lattice2017}

\end{document}